\documentclass{emulateapj}

\usepackage{epstopdf}
\usepackage{amsmath}

\newcommand{\tableskip}{\\[-6pt]}

\catcode`~=11 
\newcommand{\urltilde}{\kern -.15em\lower .7ex\hbox{~}\kern .04em}
\catcode`~=13 

\shorttitle{Effect of model-dependent covariance matrix for studying Baryon Acoustic Oscillations}
\shortauthors{Labatie et al.}

\begin{document}


\title{Effect of model-dependent covariance matrix for studying Baryon Acoustic Oscillations}


\author{A. Labatie \altaffilmark{1} and J.L. Starck}
\affil{Laboratoire AIM (UMR 7158), CEA/DSM-CNRS-Universit\'e Paris Diderot, IRFU, SEDI-SAP, Service d'Astrophysique, 
   	Centre de Saclay, F-91191 Gif-Sur-Yvette cedex, France}
\email{antoine.labatie@cea.fr}
\author{M. Lachi\`eze-Rey}
\affil{Astroparticule et Cosmologie (APC), CNRS-UMR 7164, Universit\'e Paris 7 Denis Diderot, 
            10, rue Alice Domon et L\'eonie Duquet F-75205 Paris Cedex 13, France}



\begin{abstract}
Large-scale structures in the Universe are a powerful tool to test cosmological models and constrain cosmological parameters. A particular feature of interest comes from Baryon Acoustic Oscillations (BAOs), which are sound waves traveling in the hot plasma of the early Universe that stopped at the recombination time.  This feature can be observed as a localized bump in the correlation function at the scale of the sound horizon $r_s$. As such, it provides a standard ruler and a lot of constraining power in the correlation function analysis of galaxy surveys. Moreover the detection of BAOs at the expected scale gives a strong support to cosmological models. Both of these studies (BAO detection and parameter constraints) rely on a statistical modeling of the measured correlation function $\hat{\xi}$. Usually $\hat{\xi}$ is assumed to be gaussian, with a mean $\xi_\theta$ depending on the cosmological model and a covariance matrix $C$ generally approximated as a constant (i.e. independent of the model). In this article we study whether a realistic model-dependent $C_\theta$ changes the results of cosmological parameter constraints compared to the approximation of a constant covariance matrix $C$. For this purpose, we use a new procedure to generate lognormal realizations of the Luminous Red Galaxies sample of the Sloan Digital Sky Survey Data Release 7 to obtain a model-dependent $C_\theta$ in a reasonable time. The approximation of $C_\theta$ as a constant creates small changes in the cosmological parameter constraints on our sample. We quantify this modeling error using a lot of simulations and find that it only has a marginal influence on cosmological parameter constraints for current and next-generation galaxy surveys. It can be approximately taken into account by extending the $1\sigma$ intervals by a factor $\approx 1.3$. 
\end{abstract}





\section{Introduction} 

One of the most important question in modern cosmology is to understand the nature of dark energy. This mysterious form of energy is responsible for the accelerate expansion of the Universe, and seems to account for more than 70$\%$ of the energy content of the Universe (see e.g. \cite{Kom09,Ama10,Bla11b}).

The acceleration of the expansion of the Universe was first measured with high-redshift supernovae \citep{Rie98,Per99}. The principle is to use Type Ia supernovae as standard candles in order to probe the redshift-distance relation. The same principle has been used more recently in the study of galaxy clustering at low redshift using Baryon Acoustic Oscillations (BAOs, \cite{Bas10}). These structures are remnants of acoustic waves which travelled in the plasma before recombination, when baryons and photons were coupled together. Their absolute size is given by the sound horizon scale at the baryon drag epoch, and is well constrained by measurements of the Cosmic Microwave Background (CMB), $r_s=153.3\pm 2$ Mpc \citep{Kom09}. Thus they can be used as a standard ruler to probe the redshift-distance relation.

BAOs are a very promising cosmological probe because they are less affected by systematics than other methods \citep{Alb06}. They can also be very useful to cross-check results from other probes. This has been done for example in  \cite{Bla11b}, where the combination of the WiggleZ, Sloan Digital Sky Survey (SDSS) and 6-degree Field (6dF) surveys have been used to cross-check supernovae results. As future experiments will provide more precise information, it will be critical to correctly analyze and combine these different probes. In particular one might face new challenges to deal with systematic effects that were under statistical uncertainty in previous experiments and that become important. 

Possible systematics can come from incorrect statistical modeling of the data. For example in the case of BAOs in large-scale clustering, a classical procedure is to measure the correlation function $\hat{\xi}$ and fit it to an expected correlation function $\xi_\theta$ with a dependence on cosmological parameters $\theta$. More precisely, one assumes a statistical model for $\hat{\xi}$ as a function of $\theta$ in order to compute the likelihood $\mathcal{L}_\theta(\hat{\xi})$. A common statistical model is to consider that $\hat{\xi}$ is simply gaussian, centered on the expected correlation $\xi_\theta$ and with a constant covariance matrix $C$ (i.e. independent of $\theta$).

The Gaussianity has been shown to be well verified, e.g. in \cite{Lab12b} and \cite{Man12}. However the approximation of a constant covariance $C$ has not been well studied, probably because it is very difficult to estimate a model-dependent covariance matrix $C_\theta$. Indeed the usual procedure to estimate a covariance matrix is to use a large number of realistic mock catalogues and compute the empirical covariance matrix
\begin{eqnarray}
\label{empirical_covmatrix1}
C_{ij} & = &\frac{1}{N-1} \sum_{k=1}^N [\hat{\xi}_k(r_i)-\bar{\xi}(r_i)] [\hat{\xi}_k(r_j)-\bar{\xi}(r_j)] \\
\label{empirical_covmatrix2}
\bar{\xi} & = & \frac{1}{N} \sum_{k=1}^N \hat{\xi}_k
\end{eqnarray}

Having a good estimate of the covariance matrix requires a lot of simulations. This procedure can already be long for one value of $\theta$, and it seems infeasible to apply it on a multi-dimensional grid of $\theta$ values.

As an alternative, one could find analytical formulae to estimate the covariance matrix of the correlation function $\hat{\xi}$. A recent attempt has been made in \cite{Xu12}. It starts from the analytic computation of the covariance matrix of $\hat{\xi}$ for a Gaussian density field. The covariance matrix is further modified to better match the empirical covariance matrix on mock catalogues. It is shown to reproduce the empirical covariance matrix obtained with mock catalogues, while regularizing it.

This is very interesting because it provides with little effort the covariance matrices for different input power spectra $P(k)$ of the galaxy field, i.e. a model-dependent covariance matrix $C_\theta$. However the procedure is not totally blind and requires an ad hoc fitting of the covariance matrix to mock catalogues for a given model. In particular it has not been shown that the resulting model-dependent covariance matrix $C_\theta$ is also a good estimate for other models than the one used for the fitting.

In this article we do not study this question of analytically modeling the covariance matrix. Instead we study whether this modeling is actually required, i.e. if the model-dependence of $C_\theta$ affects the statistical analysis (e.g. by changing confidence regions). We will restrict to cosmological parameter constraints using the correlation function (we will not look at the question of BAO detection for reasons explained in section \ref{cf_estimation}).

For our analysis to be feasible, we will only consider 3 parameters in $\theta$ that have the most impact on the expected correlation function $\xi_\theta$. The first parameter is the matter density $\omega_m=\Omega_m h^ 2$ which determines the horizon scale at the matter-radiation equality ($\propto \omega_m^{-1}$). It also has a little influence on the sound horizon scale ($\propto \omega_m^{-0.25}\omega_b^{-0.08}$ with $\omega_b=\Omega_b h^ 2$ the baryon density) and changes the amplitude of the BAO peak (for a constant $\omega_b$). The second parameter is $\alpha$, that determines how the correlation function is dilated when using a fiducial cosmology instead of the true cosmology to convert redshifts into distance. This parameter is the one that really probes the distance-redshift relation and it is mostly constrained by the position of the BAO peak.  Finally the third parameter is a constant bias $B=b^2$ in the correlation function that accounts for different amplitude effects (linear redshift distortions, linear galaxy bias, amplitude of matter fluctuation $\sigma_8$).

As we will estimate the covariance matrices using mock catalogues, a parameterization of $C_\theta$ with a 3-dimensional parameter $\theta=(\omega_m,\alpha,B)$ may already seem infeasible. However we will show how to optimize our simulations and the computation of the correlation function in order to make it feasible. We will show that there is in fact only 1 parameter that needs to be varied, and that the 2 other parameters can be taken into account without adding much effort.

The plan of this paper is as follows: we start in section \ref{data_section} by describing the SDSS DR7-Full data catalogue that we use. In section \ref{cf_section} we discuss the correlation function modeling and estimation. Section \ref{simulations_section} presents our new procedure to estimate a model-dependent covariance matrix $C_\theta$ with a 3-dimensional parameter $\theta=(\omega_m,\alpha,B)$ in a reasonable time. In section \ref{stat_modeling} we give results on the statistical modeling of the correlation estimator $\hat{\xi}$: absence of bias in $\hat{\xi}$, Gaussianity of $\hat{\xi}$, dependence of the covariance matrix $C_\theta$ on $\omega_m$, $\alpha$ and $B$. Finally in section \ref{constraints} we study the modeling error in parameter constraints due to the approximation of $C_\theta$ as a constant $C$. We study this modeling error on the SDSS DR7-Full $\hat{\xi}$ and we perform a quantitative analysis using simulations.

\section{Data catalogue} 
\label{data_section}

\begin{figure}[htbp]
\plotone{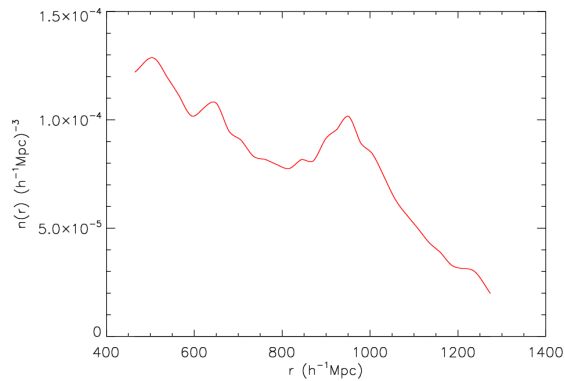} 
\caption{Observed density of the sample DR7-Full when using a flat $\Lambda$CDM fiducial cosmology with $\Omega_m=0.25$ to convert redshifts into distances.} 
\label{dens} 
\end{figure}

In this study we use the Luminous Red Galaxies sample (LRG) sample of the last Data Release 7 (DR7) of the SDSS. LRGs are selected using the algorithm in \cite{Eis01} which consists in different luminosity and color cuts using the five passbands $u,g,r,i$ and $z$. These galaxies are very luminous and good tracers of massive dark matter haloes. The sample is quasi-volume-limited (i.e. nearly of constant density) up to redshift $z \approx 0.36$ and extends up to $z \approx 0.47$ in a flux-limited way. In order to convert redshifts into distances we use a flat $\Lambda$CDM fiducial cosmology with $\Omega_m=0.25$. We plot the resulting density of the catalogue in figure \ref{dens}.

We use the DR7-Full sample of the analysis in \cite{Kaz10} that is available online\footnote{{\tt http://cosmo.nyu.edu/$\urltilde$eak306/SDSS-LRG.html}} and has the characteristics given in table \ref{sample_characteristics}.
\begin{table}[htbp]
\caption{\label{sample_characteristics}}
\begin{center}
\begin{tabular}{ll} 
\tableskip\tableline\tableline\tableskip
			      \# of LRGs &  105,831 \\
			      $z_\text{min}$ &   0.16 \\
			      $z_\text{max}$ &   0.47 \\
			      $\langle z \rangle$ &  0.324 \\
			      $M_{g,\text{min}}$ &  -23.2 \\
			      $M_{g,\text{max}}$ &  -21.2 \\
			      $\langle M_g \rangle$ &  -21.72 \\
			      Area ($\text{deg}^2$) &  7,908 \\
			      Volume ($h^{-3} \text{Gpc}^3$) & 1.58 \\
			      Density  ($10^{-5} h^{3} \text{Mpc}^{-3}$) &   6.70 \\
\tableskip\tableline\tableline\tableskip
\end{tabular}
\end{center}
NOTES.---%
Characteristics of the SDSS LRG sample used DR7-Full from \cite{Kaz10}. Volume and density have been computed with a flat $\Lambda$CDM fiducial cosmology with $\Omega_m=0.25$.
\end{table}

\begin{figure}[htbp]
\plotone{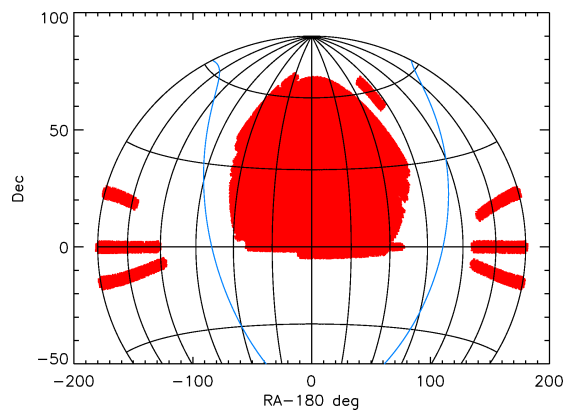} 
\caption{SDSS DR7-Full sample sky coverage in Aitoff projection. The solid blue line represents the Galactic plane which separates the Northern Contiguous region and the Southern region.} 
\label{aitoff} 
\end{figure}

The sample is mostly contiguous, with only 9.8\% outside of the main part of the Northern Galactic Cap. The number of LRGs is equal to 96763  in the Northern Galactic Cap and 9068 in the Southern Galactic Cap. We show the footprint of the survey in figure \ref{aitoff} with the Northern contiguous part and the few stripes in the Southern part (the blue line represents the Galactic plane).

\section{Correlation function modeling and estimation} 
\label{cf_section}

\subsection{Correlation function modeling}
\label{cf_modeling}
The correlation function is a second order statistic that measures the clustering of a continuous continuous field or a point process. For the galaxy field, it measures the excess of probability to find a pair of galaxies in volumes $\mathrm{d}V_1$ and $\mathrm{d}V_2$ separated by ${\bf x}$ compared to a random unclustered distribution
\begin{equation}
\mathrm{d}P_{12}=\bar{n} [1+\xi({\bf x})] \mathrm{d}V_1 \mathrm{d}V_2
\end{equation}

with $\bar{n}$ the mean density of points. Due to the cosmological principle the correlation function $\xi({\bf r})$ is isotropic so that it only depends on the norm of the separation vector $r= \| {\bf r} \|$. However we do not exactly measure the correlation function for two reasons
\begin{itemize}
\item We observe galaxies in redshift space so that there are redshift distortions in the line of sight direction
\item The choice of fiducial cosmology dilates the galaxy survey differently in the line of sight and transverse directions 
\end{itemize}
As explained later the second effect can be neglected, i.e. we can model the effect of a wrong fiducial cosmology by a single dilation factor $\alpha$ in all directions. We still want to measure the correlation function as a function of $r= \| {\bf r} \|$, so we will consider the monopole in redshift space that we denote $\xi(r)$ (and still refer to it as the correlation function as it is done in most studies)
\begin{equation}
\xi(r)=\frac{1}{4 \pi} \int \xi({\bf r}) \mathrm{d}\Omega
\end{equation}

In the plane parallel approximation and in the linear regime on large scales, the monopole correlation function in redshift space is linked to the correlation function in real space by a constant multiplicative factor independent of scale \citep{Kai86}.

When considering CDM models, the linear power spectrum can be computed up to an amplitude factor for given matter density $\omega_m$, baryon density $\omega_b$ and spectral tilt $n_s$. In our analysis we neglect the effect of $\omega_b$ and $n_s$ because they are well constrained by WMAP data \citep{Kom09}. We fix them at the maximum likelihood values of WMAP7, $\omega_b=2.227\times 10^{-2}$ and $n_s=0.966$ (we will also fix the parameter $\sigma_8=0.81$ for normalizing the linear power spectrum). So the only parameter of the linear power spectrum that we vary is the matter density $\omega_m$.

A prominent feature of the linear correlation function is the BAO peak at scale $\approx 150$ Mpc, which is due to sound waves traveling in the hot plasma before recombination, when photons are baryons were coupled together. Note however that the BAO peak is not the only effect of baryons in the linear correlation function, and that they also suppress the amplitude of fluctuations on small and intermediate scales.

Then we have to take into account the non-linear effects in the galaxy field. The first effect is due to the non-linear evolution of the matter density field, where recent advances in modeling have been made using Renormalized Perturbation Theory (\cite{Cro06}, RPT). Using RPT, it has been shown in \cite{San08} that one can have an excellent description of the correlation function for the range of scales $60h^{-1}\text{Mpc} < r< 180h^{-1}\text{Mpc}$.

In this study we use a simple model for the non-linear evolution of the matter density field. We use the HALOFIT procedure \citep{Smi03}, which provides corrections for scale-free power spectra using $N$-body simulations. Because these simulations do not include the BAO feature we also have to correct for the non-linear degradation of the acoustic peak. \cite{Eis07} found that it is well approximated by a Gaussian smoothing of the acoustic feature both in redshift and in real space. 

The power spectrum with degraded peak $P_{damped,L}$ is obtained using the linear power spectrum $P_L$ and the linear 'no wiggles' power spectrum of \cite{Eis98}, $P_{no wig,L}$
\begin{small}
\begin{equation}
P_{damped,L}(k)=P_{no wig,L}(k)+e^{-a^2 k^2/2} [P_L(k)-P_{no wig,L}(k)]
\label{P_damped_L}
\end{equation}
\end{small}
To take into account the scale-free non-linear effect, we apply to the damped power spectrum the same non-linear correction as the scale-free power spectrum $P_{no wig,L}(k)$
\begin{equation}
P_{damped,NL}(k)=\frac{P_{NL,no wig}(k)}{P_{L,no wig}(k)} P_{damped,L}(k)
\label{P_damped_NL}
\end{equation}

where $P_{NL,no wig}(k)$ is computed from $P_{L,no wig}(k)$ using the HALOFIT formula in \cite{Smi03}. We compute these power spectra using the iCosmo IDL library \citep{Ref11}.

There remains to set the value of $a$ in formula (\ref{P_damped_L}) and model the scale-dependent galaxy bias with respect to the matter density field. For these purposes we use the Large Suite of Dark Matter Simulations (LasDamas, McBride et al. 2012, in prep.). These simulations are designed to model the clustering of the SDSS DR7 for galaxies in a wide luminosity range. Galaxies are artificially placed in dark matter halos using a halo occupation distribution (HOD; \cite{Ber02}) with parameters set to match observations on the SDSS sample. 

We use the gamma release of the Las Damas simulations and more precisely the Oriana simulations that are publicly available\footnote{{\tt http://lss.phy.vanderbilt.edu/lasdamas/mocks/}}. They are composed of 40 $N$-body simulations, where each simulation can reproduce two times the 'North+South' SDSS footprint for a total of 80 realizations. Each $N$-body simulation contains $1280^3$ particles of mass $45.73 \times \, 10^{10} h^{-1}M_\odot$ with a softening parameter of $53\, h^{-1}$kpc. The cosmological parameters of the simulations are $\Omega_m=0.25$, $\Omega_\Lambda=0.75$, $\Omega_b=0.04$, $h=0.7$, $\sigma_8=0.8$ and $n_s=1$.

We use catalogues composed of LRG galaxies with $M_g < -21.2$ and $M_g>-23.2$ as the DR7-Full sample. As it is nearly volume-limited, the redshift range ($0.16<z<0.36$) is smaller than that of the DR7-Full sample. However because of a non-evolving HOD model to populate dark matter halos, the galaxy number density $n(z)$ is slowly decreasing. To address this, we compute the correlation using the random catalogue provided by the Las Damas team, which has the the same decreasing trend in its density.

We compute the correlation function using the Landy-Szalay estimator of formula (\ref{LS}). We average the measured correlation function over the 80 realizations so that we get a very good approximation of the real correlation function. On the other hand, we compute the power spectrum as in formula (\ref{P_damped_NL}) using the Las Damas cosmological parameters. We apply the Hankel transform to this power spectrum in order to obtain the corresponding correlation function. First we adjust the parameter $a$ of equation (\ref{P_damped_L}) to reproduce the non-linear degradation in the simulations and we find that the value $a=9.5 h^{-1}$Mpc gives a good result. Finally we adjust the scale-dependent galaxy bias $B(r)$ on small scales by dividing the Las Damas correlation by our model. We find a scale-dependent correction of $\approx 10\%$ at $r=5h^{-1}$Mpc which slowly decreases up to $r=55 h^{-1}$Mpc.

We thus obtain the galaxy correlation function
\begin{equation}
\xi_{galaxy,\omega_m}(r)=B(r) \, \xi_{damped,NL}(r)
\end{equation}
where $\xi_{damped,NL}(r)$ is obtained by the Hankel transform of $P_{damped,NL}(k)$ of formula (\ref{P_damped_NL}) with the choice $a=9.5 h^{-1}$Mpc in equation (\ref{P_damped_L}). We keep $B(r)$ and $a$ fixed in our analysis, so that $\xi_{galaxy,\omega_m}$ only depends on the linear power spectra $P_L$ and $P_{no wig,L}$ of equation (\ref{P_damped_L}). And as we already explained, we only vary the parameter $\omega_m$ in the linear power spectra. So the correlation function $\xi_{galaxy,\omega_m}$ only has a dependence on $\omega_m$.

We introduce two additional parameters in the model correlation function. The first parameter $\alpha$ accounts for a dilation of the galaxy survey due to an incorrect choice of fiducial cosmology to convert redshifts into distances. This parameter is actually the one that is probed by the localization of the BAO peak and the standard ruler property. It was shown that a wrong choice of fiducial cosmology approximately translates into a dilation of the galaxy survey and thus of the correlation function \citep{Eis05,Pad08} by a factor $\alpha=D_V(z_{eff}) /D_{V,fid}(z_{eff})$ with $z_{eff}=0.3$ the effective redshift of our sample, and $D_V(z)$ the 'dilation scale' at redshift $z$
\begin{equation}
D_V(z)= \left[ D_M(z)^2 \frac{cz}{H(z)} \right]^{1/3}
\end{equation}

where $H(z)$ is the Hubble parameter and $D_M(z)$ is the comoving angular diameter distance at redshift $z$. Our choice of a flat $\Lambda$CDM fiducial cosmology with $\Omega_m=0.25$ gives $D_{V,fid}(z_{eff}=0.3)=1180$ Mpc.

Next we introduce a constant amplitude factor $b$ to model variations of $\sigma_8$, linear redshift distortions and linear galaxy bias. So we obtain the final model correlation function as a function of $\omega_m, \alpha$ and $B=b^2$
\begin{equation}
\xi_{\omega_m,\alpha,B}(r)=b^2 \xi_{galaxy,\omega_m}(\alpha \, r)
\end{equation}

Finally we bin the model correlation function equivalently as when it is estimated by pair counting, i.e. for a bin $[r_i -\mathrm{d}r/2,r_i +\mathrm{d}r/2]$
\begin{equation}
\xi_{\omega_m,\alpha,B}(r_i)= \frac{\int_{r_i -\mathrm{d}r/2}^{r_i +\mathrm{d}r/2}  \xi_{\omega_m h^2,\alpha,B}(r) \, r^2 \,  \mathrm{d}r } {\int_{r_i-\mathrm{d}r/2}^{r_i+\mathrm{d}r/2}  \, r^2 \,  \mathrm{d}r }
\end{equation}
In all this study we use a $\mathrm{d}r=10 h^{-1}$Mpc binning from $20 h^{-1}$Mpc  to $200 h^{-1}$Mpc corresponding to $n=18$ bins.

\subsection{Correlation function estimation}
\label{cf_estimation}
Most estimators of the correlation function use random unclustered catalogues (i.e. Poisson catalogues with no correlation) and compare the excess of pairs of data points separated by a distance $r$ compared to pairs of random points. Different estimators have been proposed and compared \citep{Pon99,Lab12a}. The recommendation is to use either the Hamilton estimator \citep{Ham93} or the Landy-Szalay estimator \citep{Lan93}. They have been shown in \cite{Lab12a} to have lower variance than the other estimators and negligible bias for current galaxy surveys. Most studies are using the Landy-Szalay estimator, and we will also use it here. It is given by
\begin{equation}
\hat{\xi}(r) =  1 + {N_{RR} \over N_{DD}} {  DD(r) \over  RR(r)}  -  2 {N_{RR} \over N_{DR}}  { DR(r) \over RR(r)}
\label{LS}
\end{equation}
with $DD(r)$, $RR(r)$, $DR(r)$ the number of pairs at a distance in $[r \pm \mathrm{d}r/2]$ of respectively data-data, random-random, data-random points and $N_{DD}$, $N_{RR}$, $N_{DR}$ the total number of corresponding pairs in the catalogues.

Formula (\ref{LS}) corresponds to the case where all galaxies are weighted equally in the estimator. This is optimal for volume-limited surveys but it is not optimal when the galaxy mean density depends on redshift. An approximately optimal weighting, which depends on the distance $r$ at which we estimate the correlation function, is given in \cite{Ham93} by
\begin{equation}
w_i=\frac{1}{1+\bar{n}\Phi_i J(r)}
\label{optimal_weight}
\end{equation}
where $\Phi_i$ is the selection function at the position of the galaxy $i$, $\bar{n}$ is the expected density of the catalogue before the selection function is applied and $J(r)$ is the integral of the real correlation function
\begin{equation}
J(r)=\int_{V_r} \xi({\bf s}) \mathrm{d}^3 {\bf s}=4\pi \int^r_0 \xi(s) s^2 \mathrm{d}s
\end{equation}

There is still a constraint not to introduce a bias, which is that the weighted density of the random catalogue and data catalogue must be proportional (i.e. there can only be a multiplicative factor of difference between the two). When introducing weights as in formula (\ref{optimal_weight}) the pair-counting quantities ($DD, RR, DR$) are modified in the Landy-Szalay estimator of equation (\ref{LS}). Instead of adding +1 for each pair, we simply add $w_i w_j$, with $w_i$ and $w_j$ the weights of each point of the pair.

When computing the correlation function of the DR7-Full sample we do not try to apply such optimal weights. We only take care of the fiber collision problem which locally changes the density of galaxies. We apply the same weights as in \cite{Kaz10}, that upweight groups of galaxies which are close enough to be affected by fiber collisions. Concerning the angular incompleteness and the varying density with redshift, they are taken into account in the random catalogue. So overall the weighted density in the data and random catalogues are proportional.

We use the same random catalogue as in \cite{Kaz10} which is also available online\footnote{{\tt http://cosmo.nyu.edu/$\urltilde$eak306/SDSS-LRG.html}} . It is composed of $\approx 1.66$ million points, i.e. $\approx 16$ times the number of galaxies in the data. 

\begin{figure}  
\plotone{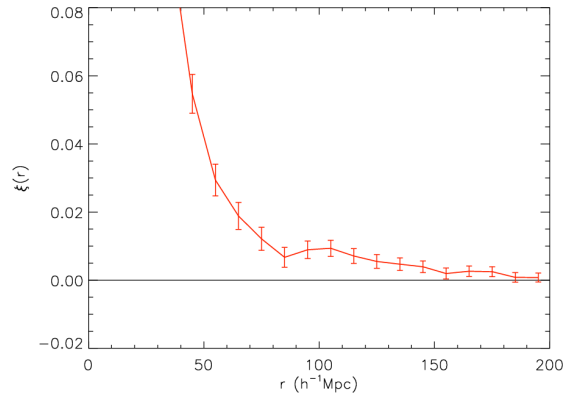} 
\caption{Estimated correlation function of the SDSS DR7-Full sample $\hat{\xi}$ with a flat $\Lambda$CDM fiducial cosmology with $\Omega_m=0.25$. We give the error bars as the diagonal part $\sqrt{C_{ii}}$ of the covariance matrix obtained from 2000 lognormal simulations with parameters $\omega_m=0.13$, $\alpha=1$ and $b=2.5$. The BAO peak is a bit wider than expected, which is explained by signal variance in \cite{Kaz10}.} 
\label{correlation_dr7} 
\end{figure}

We plot in figure \ref{correlation_dr7} the measured correlation function of the data sample, with a BAO peak a bit wider than expected. This was also found in \cite{Mar09} on a SDSS DR7 LRG volume-limited sample. Yet the study \cite{Kaz10} concludes that this is not due to systematics but only to signal variance. Note also that the BAO reconstruction technique used in \cite{Pad12} on the same sample leads to a sharpening of the BAO peak. However, without applying this technique or introducing nuisance parameters, the wide BAO peak results in a low BAO detection level and also a shift towards values $\alpha <1$ (see section \ref{constraints}).

A lot of studies on the clustering of the SDSS DR7 LRG sample focused only on the position of the BAO peak. This is done either by using peak finding techniques as in \cite{Kaz10}, or by introducing nuisance parameters for the global shape of the correlation function (or power spectrum) which are marginalized over (e.g. spline functions in \cite{Per10} or inverse polynomials in \cite{Xu12}). 

In the latter case, this enables to obtain high BAO detection levels, that we do not manage to obtain here otherwise ($3.6\sigma$ in \cite{Per10}  and $3\sigma$ before reconstruction in \cite{Xu12}). Therefore we will not study the BAO detection here. Another reason is that the presence of BAOs in large-scale structures is becoming hard to dispute after recent results from the surveys WiggleZ ($3.2\sigma$ detection in \cite{Bla11a}), 6dF ($2.4\sigma$ detection in \cite{Beu11}) and BOSS (5$\sigma$ detection in \cite{And12}). Finally let us mention that wavelet analysis also enabled to obtain high level of detection using SDSS DR7 samples ($4.4\sigma$ in \cite{Arn12} and $4\sigma$ in \cite{Tia11}).

So we will focus on cosmological parameter constraints using the SDSS DR7-Full sample described in section \ref{data_section}. Because we use a relatively simple correlation function modeling, our study is not meant to improve cosmological parameter constraints. We only attempt to quantify the modeling error introduced by the approximation of a constant covariance $C$ instead of a model-dependent $C_\theta$.

\section{Lognormal simulations} 
\label{simulations_section}

In this section we describe our procedure for generating lognormal simulations that will provide us with a model-dependent covariance matrix $C_\theta$. In our lognormal simulations we use the same sky coverage and the same number density as in the SDSS DR7-Full sample. 

To generate lognormal realizations we use the same method as in \cite{Lab12a}: we generate a continuous galaxy field in a cube from an input correlation function $\xi_\theta$, we apply the SDSS DR7-Full selection function (which incorporates the angular mask and the number density), and finally we Poisson sample the resulting continuous field.

For computational reasons we do not estimate the correlation function $\hat{\xi}$ on the full sky, but separately on the Northern Galactic Cap, $\hat{\xi}_{NGC}$ and Southern Galactic Cap, $\hat{\xi}_{SGC}$, which can be considered as independent. Also for computational reasons we use random catalogues with the same density as the SDSS DR7-Full sample.

From these measurements we obtain the model-dependent covariance matrices $C_{NGC,\theta}$, and $C_{SGC,\theta}$ by computing the empirical covariance matrices (as in equations (\ref{empirical_covmatrix1}) and (\ref{empirical_covmatrix2})). For each simulation, corresponding to a parameter $\theta$, we obtain the full correlation function $\hat{\xi}$ by the same optimal linear combination as in \cite{Whi11} (see appendix \ref{optimal_combination})
\begin{eqnarray} 
\hat{\xi} & = &C_{\theta} \left[ C^{-1}_{NGC,\theta} \hat{\xi}_{NGC}+ C^{-1}_{SGC,\theta} \hat{\xi}_{SGC} \right] \\
C_\theta & = & \left( C^{-1}_{NGC,\theta} +C^{-1}_{SGC,\theta} \right)^{-1}
\end{eqnarray}
with $C_\theta$ the resulting covariance matrix of the full correlation $\hat{\xi}$.

As we stated in section \ref{cf_modeling} we only take into account 3 main parameters in the correlation function, i.e. $\theta= (\omega_m,\alpha,B)$. 

The parameter $\omega_m$ changes the whole shape of the correlation function, so we have no choice but to generate different sets of lognormal simulations for different values of $\omega_m$. We choose to use 5 values $\omega_m=0.08,0.105,0.13,0.155,0.18$ and simply interpolate linearly the covariance matrix for intermediate values (more precisely, each coefficient of the covariance matrix is linearly interpolated). 

The parameter $\alpha$, on the other hand, only creates a dilation of the galaxy survey and thus of the apparent correlation function. This is only a geometrical effect due to a wrong fiducial cosmology. It is thus possible to take it into account using  a single set of simulations. 

First we must take into account that if the survey extends from a minimum distance $r_{min}$ to a maximum distance $r_{max}$ in fiducial coordinates, it extends from $\alpha \, r_{min}$ to $\alpha \, r_{max}$ in comoving coordinates. So for a simulation parameter $\alpha$, one must consider cuts at these distances $\alpha \, r_{min}$ and $\alpha \, r_{max}$ and then artificially dilate the survey by a factor $\alpha$ to mimic the effect of a wrong fiducial cosmology.

So instead of producing simulations that extend from $r_{min}$ to $r_{max}$, we produce simulations that extend from $\alpha_{min} \, r_{min}$ to $\alpha_{max} \, r_{max}$, where $\alpha_{min}$ and $\alpha_{max}$ are the minimum and maximum values of $\alpha$ considered. In this way we are always able to consider cuts at distances $\alpha \, r_{min}$ and $\alpha \, r_{max}$. In this study we choose $\alpha_{min}=0.8$ and $\alpha_{max}=1.2$. Given the value $D_{V,fid}(0.3)=1180$ Mpc for our fiducial cosmology, we get a probed range $D_V(0.3)\in [944 \mbox{ Mpc}, 1416 \mbox{ Mpc}]$.

There is another complication because the apparent density must be in agreement with the one observed in the data catalogue. So in addition to the cuts between $\alpha \, r_{min}$ and $\alpha \, r_{max}$, we introduce a varying selection function that depends on $\alpha$, so that the observed density after the dilation by $\alpha$ agrees with the one of the data catalogue. 

We developed an optimized procedure for computing the correlation function in this context. First, because the correlation function is estimated by pair-counting, the estimation can be done in comoving coordinates (i.e. before the dilation) and the dilation is only applied after the pair-counting by dilating bin ranges. The density in comoving space is given by
\begin{equation}
n_\alpha(r)=\frac{1}{\alpha^3} n\left(\frac{r}{\alpha}\right)
\end{equation}
with $n(r)$ the observed density in the data catalogue and the factor $1/\alpha^3$ accounting for the change of density because of the dilation. 

So the original lognormal simulations are generated with a density $n_{max}(r)=\max_\alpha n_\alpha(r)$. Let us define the selection function $\Phi_\alpha(r)=n_\alpha(r)/n_{max}(r)$. We apply this selection function for every value of $\alpha$ in the following way: for each galaxy at distance $r$ in the original simulation, we generate a random uniform variable $u \in [0,1]$. Then the galaxy belongs to the simulation with value $\alpha$ if $u<\Phi_\alpha(r)$. 

For each galaxy ${ \bf x_i}$ we end up with a sequence of intervals $[\alpha_i,\alpha'_i]$ for which the galaxy belongs to the simulations. To optimize the computation of the correlation function we create a new galaxy at the same position for every distinct interval $[\alpha_i,\alpha_{i+1}]$. 

Let us consider only the pair counting term $DD$, with the same argument that could be applied for $DR$ and $RR$. For every $r$ we consider an array $(DD_{\alpha_i,raw}(r))_{i=1,\dots,n}$ corresponding to the grid $\alpha=(\alpha_1,\dots,\alpha_n)$. This counts the number of pairs to add from $DD_{\alpha_i}(r)$ to obtain $DD_{\alpha_{i+1}}(r)$. 

For every pair $({\bf x_k},{\bf x_l})$ with $\alpha$ ranges respectively equal to $[\alpha_k,\alpha_{k'}]$ and $[\alpha_l,\alpha_{l'}]$, the pair belongs to the simulations for the range $[\max (\alpha_k,\alpha_l), \min (\alpha_{k'},\alpha_{l'})]=[\alpha_{\max(k,l)},\alpha_{\min(k',l')}]$. So we add +1 to $DD_{\alpha,raw}(r)$ for $\alpha=\alpha_{\max(k,l)}$ and add -1 for $\alpha=\alpha_{\min(k',l')+1}$. In the end we obtain the $\alpha$ dependent $DD_\alpha(r)$ as 
\begin{equation}
DD_{\alpha_i}(r)=\sum_{j=0}^i DD_{\alpha_j,raw}(r)
\end{equation}

Finally we only have to perform the dilation on $DD_{\alpha_i}(r)$ that was computed in comoving coordinates
\begin{equation}
DD^{final}_{\alpha}(r)=DD_{\alpha}(\alpha \, r)
\end{equation}

This whole procedure enables to obtain $DD$, $DR$ and $RR$ for every $r$ and every $\alpha$ with a time increased only by a factor $\approx 4$ instead of being proportional to the number of $\alpha$ values.

Finally let us turn to the third parameter $B=b^2$, which changes the real galaxy distribution in comoving space, just like $\omega_m$. But because it is simply a constant multiplicative factor $B$ in the correlation function, it should give approximately a factor $B^2$ in the covariance matrix of $\hat{\xi}$. We recall that there are two different sources of noise in the estimator $\hat{\xi}$
\begin{itemize}
\item Cosmic variance due to the finite extent of the catalogue
\item Shot noise due to the finite number of galaxies to map an underlying continuous field
\end{itemize}

The approximation of a covariance matrix scaling as $B^2$ is valid when we can neglect the shot noise contribution compared to the cosmic variance contribution. So obviously it is better verified for large values of $b$. However we verify in section \ref{results_covmatrix} that it is a good approximation around reasonable values of $b$, with the approximation $B^2 C$ being much closer to the real covariance matrix than the approximation of a constant $C$. So this parameter will actually be treated without any need for more simulations.

Our main set of simulations will be performed with $b=2.5$ (note that this value is with respect to the real space correlation, i.e. without the boost factor of \cite{Kai86}). For each value of $(\omega_m,\alpha)$ we will use $N=2000$ lognormal simulations to estimate the covariance matrix $C_{\omega_m,\alpha}$.

\section{Results on the statistical modeling of $\hat{\xi}$} 
\label{stat_modeling}

\subsection{Absence of bias in $\hat{\xi}$}
We first test whether there is a bias affecting the estimators of the correlation function in our lognormal simulations. This is important for cosmological parameter constraints because the expected value of $\hat{\xi}$ is assumed to be from a given model $\xi_\theta$ (see section \ref{constraints})
\begin{equation}
 \exists \, \theta \in \Theta \,\, \mbox{s.t.} \,\, \hat{\xi} \thicksim \mathcal{N}\left( \xi_\theta ,C_\theta \right) \\
\end{equation} 

To verify that the bias is negligible we compute the mean of the measured correlation function for $\alpha=1$ and for the different values $\omega_m=0.08,0.105,0.13,0.155,0.18$, using $N=2000$ lognormal simulations in each case
\begin{equation}
\bar{\xi}_{\omega_m}= \frac{1}{N} \sum_{k=1}^N \hat{\xi}_{k,\omega_m}
\end{equation}

\begin{figure}  
\plotone{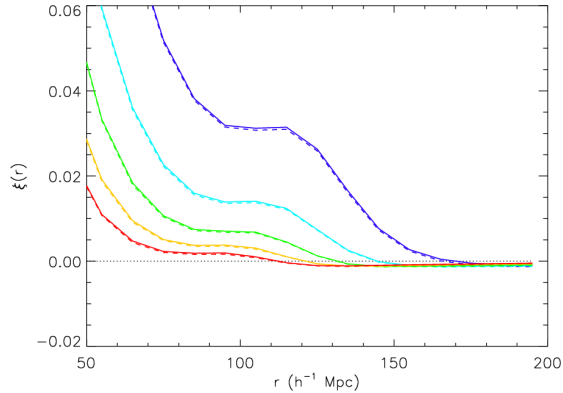} 
\caption{Mean estimators $\bar{\xi}_{\omega_m}$ in dashed lines compared to the real correlation function $\xi_{\omega_m}$ in solid lines for $\alpha=1$ and for $\omega_m=0.08$ (purple), $0.105$ (light blue), $0.13$ (green), $0.155$ (yellow), $0.18$ (red).} 
\label{bias} 
\end{figure}

We plot in figure \ref{bias} the resulting mean estimators $\bar{\xi}_{\omega_m}$ compared to the real correlation function $\xi_{\omega_m}$, which is given as the lognormal simulations input. Figure \ref{bias} shows a very good agreement, i.e. that the estimators are nearly unbiased.

\subsection{Verification of the Gaussianity of $\hat{\xi}$}
Now we want to verify the Gaussianity of the measured correlation function $\hat{\xi}$, i.e. again to verify that the following hypothesis is realistic
\begin{equation*}
 \exists \, \theta \in \Theta \,\, \mbox{s.t.} \,\, \hat{\xi} \thicksim \mathcal{N}\left( \xi_\theta ,C_\theta \right) \\
\end{equation*}

For this we use the correlation function estimates $\hat{\xi}$ on the $N=80$ Las Damas realizations presented in section \ref{cf_modeling}. Indeed they are more realistic than our lognormal realizations. For example the broadening of the BAO feature appears through non-linear evolution in the Las Damas simulations, whereas it is simply 'injected' through the input correlation function in our lognormal realizations.  

First we compute the empirical mean and empirical covariance matrix of the LasDamas realizations
\begin{eqnarray}
\bar{\xi} & = & \frac{1}{N} \sum^N_{k=1} \hat{\xi}_k \\
C_{ij} & = & \frac{1}{N-1} \sum^N_{k=1} \left[ \hat{\xi}_k (r_i)-\bar{\xi}(r_i) \right] \left[ \hat{\xi}_k(r_j)-\bar{\xi}(r_j) \right]
\end{eqnarray} 

We then compute the $\chi^2$ statistic for each realization $\hat{\xi}_k$, which should approximately follow a $\chi_n^2$ law with $n=18$ if the measurement $\hat{\xi}$ is Gaussian

\begin{eqnarray}
\chi^2 	& = &  \left \langle \hat{\xi}-\bar{\xi}, C^{-1} (\hat{\xi}-\bar{\xi}) \right \rangle \\
 	 	& = & \sum_{1\leq i,j\leq n}  \left[ \hat{\xi}(r_i)-\bar{\xi}(r_i) \right] C^{-1}_{i,j}  \left[ \hat{\xi}(r_j)-\bar{\xi}(r_j) \right]
\end{eqnarray}

\begin{figure}  
\plotone{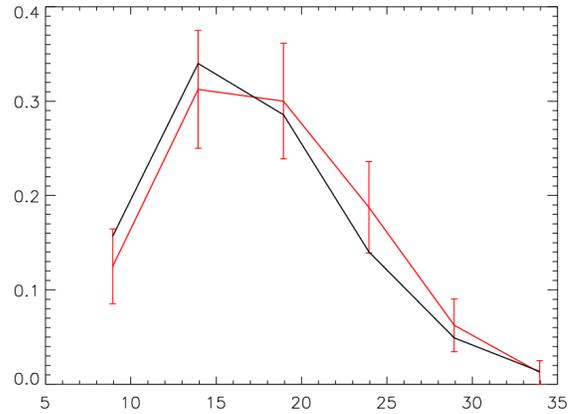} 
\caption{Estimated pdf of $\chi^2$ (red) using the histogram on the 80 Las Damas realizations and pdf of a $\chi^2_{18}$ distribution (black). Error bars give the Poisson uncertainty in the estimate due to finite number of realizations.} 
\label{chi2_histo} 
\end{figure}

We show on figure \ref{chi2_histo} the histogram of $\chi^2$ on the 80 Las Damas realizations compared to the probability density function (pdf) of a $\chi^2_n$ variable with $n = 18$. We can see the very good agreement between the two distributions.

\subsection{Dependence of $C_\theta$ on $\omega_m$, $\alpha$ and $B$}
\label{results_covmatrix}
Here we describe the dependence of $C_\theta$ (obtained from our full set of lognormal simulations) with respect to $\omega_m$, $\alpha$ and $B$.

First we check that that the dependence of $C_\theta$ on $B=b^2$ can actually be approximated as $C_{\omega_m,\alpha,B} \propto B^2 C_{\omega_m,\alpha}$. For this we compare the covariance matrix $C_1=C_{\omega_m,\alpha,B_1}$ to the covariance matrix $C_2=C_{\omega_m,\alpha,B_2}$ obtained in each case from $N=2000$ lognormal simulations, respectively with $\omega_m=0.13$, $\alpha=1$, $B_1=2.5^2$ and $\omega_m=0.13$, $\alpha=1$, $B_2=3.0^2$.

We compute the L2 distance between $(B_2/B_1)^2 C_1$ and $C_2$, and compare it to the L2 distance between $C_1$ and $C_2$
\begin{equation}
\frac{\| (B_2/B_1)^2 C_1-C_2 \|_2}{\| C_1-C_2 \|_2} = 0.22
\end{equation}

So we obtain that the approximation $C_{\omega_m,\alpha,B} \propto B^2 C_{\omega_m,\alpha}$ is 5 times better than the approximation of a constant covariance matrix, which justifies our approximation.

\begin{figure}[h]
\plotone{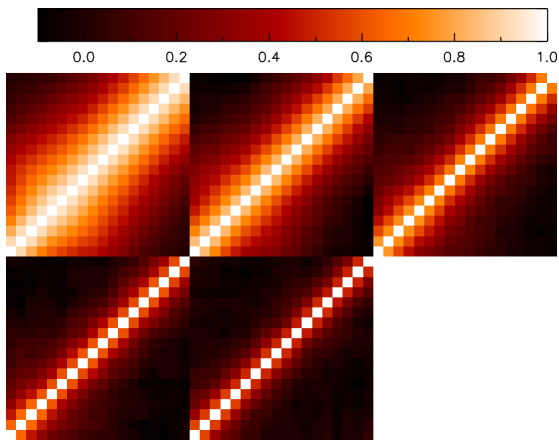} 
\caption{Dependence of $\rho_\theta$ with respect to $\omega_m$, in the case $\alpha=1$ and $B=2.5^2$. We plot $\rho_\theta$ for $\omega_m=0.08$ (top left), $0.105$ (top middle), $0.13$ (top right), $0.155$ (bottom left), $0.18$ (bottom middle). The correlation between bins strongly increases for smaller values of $\omega_m$. We have plotted the $n=18$ bins of size $\mbox{d} r = 10 h^{-1}$Mpc from $20 h^{-1}$Mpc to $200 h^{-1}$Mpc. } 
\label{omegamh2_dependence1} 
\end{figure}

\begin{figure}[h]
\plotone{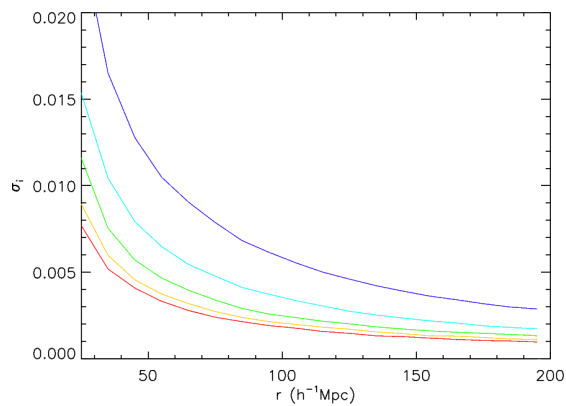} 
\caption{Dependence of $\sigma_\theta=\left(\sqrt{C_{\theta,ii}} \right)$ with respect to $\omega_m$, in the case $\alpha=1$ and $B=2.5^2$. We plot $\sigma_\theta$ for $\omega_m=0.08$ (purple), $0.105$ (light blue), $0.13$ (green), $0.155$ (yellow), $0.18$ (red). The diagonal variance strongly increases for smaller values of $\omega_m$.} 
\label{omegamh2_dependence2} 
\end{figure}

Next we outline the significant dependence of $C_\theta$ with respect to the two other parameters $\omega_m$ and $\alpha$. We start by analyzing the dependence of $C_\theta$ with respect to $\omega_m$ in the case $\alpha=1$ and $B=2.5^2$. We show on figure \ref{omegamh2_dependence1} and \ref{omegamh2_dependence2} the variations of $C_\theta$ for $\omega_m=0.08,0.105,0.13,0.155,0.18$. For clarity reasons we distinguish between the correlation matrix $\rho_\theta$ (i.e. the covariance matrix normalized by the diagonal elements) of formula (\ref{correlation_matrix}) and the diagonal part $\sigma_\theta = \left(\sqrt{C_{\theta,ii}} \right)$, which fully describe the covariance matrix together. 

\begin{equation}
\rho_{\theta,ij}=\frac{C_{\theta,ij}}{\sqrt{C_{\theta,ii} C_{\theta,jj}}}=\frac{1}{\sigma_{\theta,i} \sigma_{\theta,j}} C_{\theta,ij}
\label{correlation_matrix}
\end{equation}

We recall that the correlation function has $n=18$ bins of size $\mbox{d}r = 10h^{-1}$Mpc from $20h^{-1}$Mpc to $200 h^{-1}$Mpc. We find a strong dependence of the whole covariance matrix with respect to $\omega_m$, i.e. both the diagonal part $\sigma_\theta$ and the correlation matrix $\rho_\theta$ have a strong dependence on $\omega_m$.

\begin{figure}[h]
\plotone{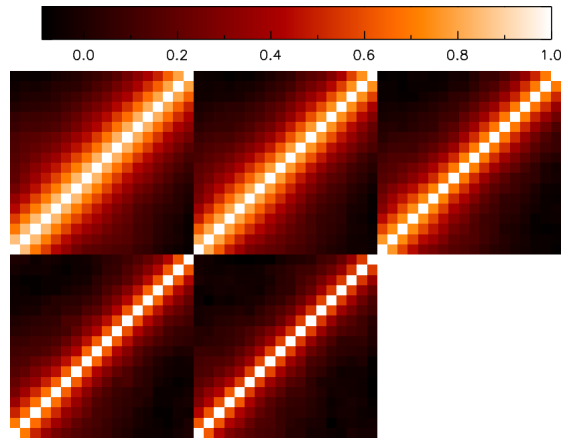} 
\caption{Dependence of $\rho_\theta$ with respect to $\alpha$, in the case $\omega_m=0.13$ and $B=2.5^2$. We plot $\rho_\theta$ for $\alpha=0.8$ (top left), $0.9$ (top middle), $1.0$ (top right), $1.1$ (bottom left), $1.2$ (bottom middle). The correlation between bins increases for smaller values of $\alpha$. We have plotted the $n=18$ bins of size $\mbox{d} r = 10 h^{-1}$Mpc from $20 h^{-1}$Mpc to $200 h^{-1}$Mpc.} 
\label{alpha_dependence1} 
\end{figure}

\begin{figure}[h]
\plotone{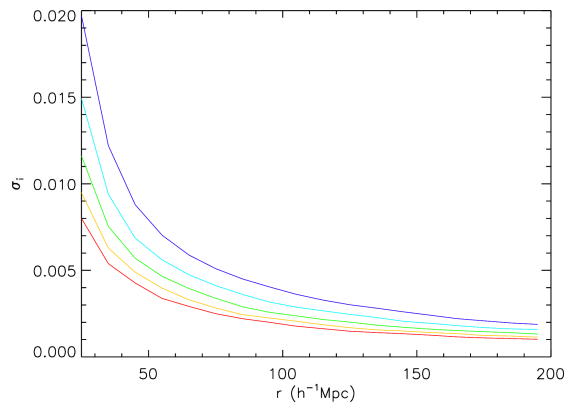} 
\caption{Dependence of $\sigma_\theta$ with respect to $\alpha$, in the case $\omega_m=0.13$ and $B=2.5^2$. We plot $\sigma_\theta$ for $\alpha=0.8$ (purple), $0.9$ (light blue), $1.0$ (green), $1.1$ (yellow), $1.2$ (red). The variance increases for smaller values of $\alpha$. }
\label{alpha_dependence2} 
\end{figure}

Finally we analyze the dependence of $C_\theta$ with respect to $\alpha$ in the case $\omega_m=0.13$ and $B=2.5^2$. We show on figure \ref{alpha_dependence1} and \ref{alpha_dependence2} the variations of $C_\theta$ for $\alpha=0.8, 0.9, 1.0, 1.1, 1.2$, again plotting separately the correlation matrix $\rho_\theta$ and the diagonal part $\sigma_\theta$. We also find a dependence of both $\rho_\theta$ and $\sigma_\theta$ with respect to $\alpha$ but this dependence is not as strong as for $\omega_m$. Note that this conclusion is dependent on the ranges of parameter values, but here we considered pretty standard ranges.

\section{Effect of $C_\theta$ for cosmological parameter constraints}
\label{constraints}
To obtain cosmological parameter constraints from BAOs one usually perform a likelihood analysis using the whole correlation function \citep{Eis05,San09,Beu11,Bla11a,Bla11b} or power spectrum \citep{Col05,Teg06,Pad07,Rei10,Ho12}, though some studies effectively restrict the analysis to the position of the BAO peak \citep{Per07,Per10,Kaz10,Meh12}. 

One supposes that the following hypothesis is true and wants to constrain the parameter $\theta$
\begin{equation*}
  \exists \, \theta \in \Theta \,\, \mbox{s.t.} \,\, \hat{\xi} \thicksim \mathcal{N}\left( \xi_\theta ,C_\theta \right) 
\end{equation*}

To obtain posterior information on $\theta$ one needs a Bayesian point of view by assuming a prior $p(\theta)$. Then the posterior of $\theta$ knowing the measurement $\hat{\xi}$ is given by the Bayes' theorem
\begin{equation}
p(\theta \, | \,\hat{\xi})  \propto  p(\theta) \, p(\hat{\xi} \, | \, \theta)=p(\theta) \, \mathcal{L}_{\theta}(\hat{\xi}) 
\end{equation}

The combination of the measurement $\hat{\xi}$ with other independent experiments can be done inside the prior. For example with CMB data the posterior is given by
\begin{eqnarray}
p(\theta \, | \, \mbox{CMB}, \hat{\xi}) & \propto & p(\theta, \mbox{CMB}, \hat{\xi} ) \nonumber  \\
 					& \propto & p(\theta, \mbox{CMB}) \, p(\hat{\xi} \, | \, \theta,\mbox{CMB}) \nonumber \\
					 & \propto & p(\theta \, | \, \mbox{CMB}) \,  \mathcal{L}_{\theta}(\hat{\xi})
\end{eqnarray}
where we used the independence of $\hat{\xi}$ and CMB measurement. Adding the CMB measurement is thus equivalent to using a prior $p(\theta)=p(\theta\, | \, \mbox{CMB})$.

To constrain $\theta$ only from the measurement $\hat{\xi}$ the question of choosing a prior $p(\theta)$ can be difficult. In this study we take a constant prior $p(\theta)$, but note that this choice is arbitrary. So the posterior is equivalent to the likelihood
\begin{equation}
 \mathcal{L}_\theta(\hat{\xi}) \propto  |C_\theta|^{-1/2} e^{ -\frac{1}{2} \left\langle \hat{\xi}-\xi_\theta, C^{-1}_\theta (\hat{\xi}-\xi_\theta) \right\rangle} 
\end{equation}

In all the following we compare the posterior obtained using our model-dependent  $C_\theta$ to the posterior obtained with constant covariance matrix $C=C_{\theta_0}$ for the particular value $\theta_0=(\omega_m, \alpha, B)=(0.13,1.0,2.5^2)$. We only plot the 2D posteriors $p(\omega_m, D_V(0.3)\, | \,\hat{\xi})$ (we recall the simple relation $\alpha=D_V(0.3)/D_{V,fid}(0.3)$), i.e. after marginalizing over $B=b^2$.
\begin{equation}
p(\omega_m, D_V(0.3)\, | \,\hat{\xi})= \int_B p(\omega_m, D_V(0.3), B \, | \,\hat{\xi}) \mathrm{d}B
\end{equation}
where we will consider the following grid: $B\in [4.0,9.0]$ with grid step $\mathrm{d}B=0.01$, $\omega_m \in [0.08,0.18]$ with grid step 0.00025 and $\alpha \in [0.8,1.2]$ with grid step 0.001. This grid in $\alpha$ corresponds to a grid $D_V(0.3) \in [944\mbox{ Mpc},1416\mbox{ Mpc}]$ with grid step 1.18 Mpc.

\subsection{Effect of $C_\theta$ on the SDSS DR7-Full $\hat{\xi}$}
\label{model_error_dr7}
Here we work with the SDSS DR7-Full estimated correlation function $\hat{\xi}$ of figure \ref{correlation_dr7}.

\begin{figure}[h]
\plotone{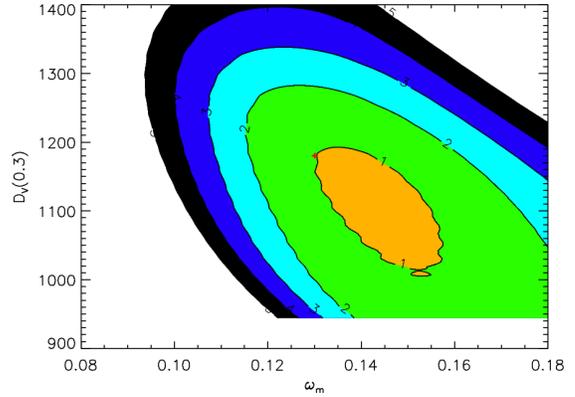} 
\caption{Posterior $p(\omega_m, D_V(0.3) \, | \, \hat{\xi})$ in the case of constant covariance matrix $C=C_{\theta_0}$, with $\theta_0=(\omega_m, \alpha,B)=(0.13,1.0,2.5^2)$ (position of the red cross on the figure), for the SDSS DR7-Full measurement $\hat{\xi}$. We plot the $1\sigma$ to $5\sigma$ confidence regions with the approximation that $p$ is a 2-dimensional Gaussian. They correspond respectively to $-2\ln(p)=-2\ln(p_{max})+2.29,6.16,11.81,19.32,28.74$ (see section 'Confidence Limits on Estimated Model Parameters' in \cite{Pre07}). We obtain the 1-dimensional constraints $\omega_m= 0.145 \pm 0.016$ (10.8\% precision) and $D_V(0.3)= 1104 \pm 105$ Mpc (9.5\% precision).} 
\label{constraints_dr7_ccov} 
\end{figure}

\begin{figure}[h]
\plotone{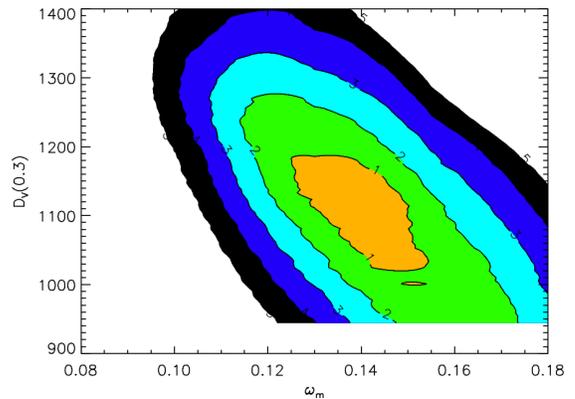} 
\caption{Posterior $p(\omega_m,D_V(0.3) \, | \, \hat{\xi})$ in the case of model-dependent covariance matrix $C_\theta$ for the SDSS DR7-Full measurement $\hat{\xi}$. We obtain the 1-dimensional constraints $\omega_m=0.140 \pm 0.011$ (7.9\% precision) and $D_V(0.3)=1114 \pm 74$ Mpc (6.7\% precision). There is a small shift in the position of the posterior's maximum and the confidence regions get smaller when considering a model-dependent $C_\theta$.} 
\label{constraints_dr7_vcov} 
\end{figure}

We plot in figures \ref{constraints_dr7_ccov}  and \ref{constraints_dr7_vcov} the posterior $p(\omega_m, D_V(0.3) \, | \, \hat{\xi})$, respectively for a constant covariance matrix $C=C_{\theta_0}$ and for a model-dependent covariance matrix $C_\theta$.

First we can notice that the posterior $p(\omega_m,D_V(0.3) \, | \, \hat{\xi})$ is less regular and more 'noisy' in the case of model-dependent $C_\theta$. This can be easily explained by the noise in the estimation of $C_\theta$. 

We also notice that the 2-dimensional posterior cannot be so well approximated by a 2-dimensional Gaussian (characterized notably by elliptical contours), especially in the case of constant $C$. We attribute this to the behavior of the model correlation function $\xi_\theta$ for high $\omega_m$ and low $\alpha$ (bottom right of figure \ref{constraints_dr7_ccov}).

From the 2-dimensional posteriors we compute 1-dimensional posteriors on $\omega_m$ and $D_V(0.3)$, by marginalizing over the other parameter. Then we compute 1-dimensional constraints, that we express as a symmetric 68\% confidence interval ($1\sigma$ interval) around the posterior's maximum.

In the case of constant covariance matrix $C$, we obtain the constraints $\omega_m= 0.145 \pm 0.016$ (10.8\% precision) and $D_V(0.3)= 1104 \pm 105$ Mpc (9.5\% precision). Whereas in the case of model-dependent covariance matrix $C_\theta$, we obtain the constraints $\omega_m=0.140 \pm 0.011$ (7.9\% precision) and $D_V(0.3)=1114 \pm 74$ Mpc (6.7\% precision). In terms of $\alpha$, this gives respectively the constraints $\alpha=0.935 \pm 0.089$ for constant $C$ and $\alpha=0.944 \pm 0.063$ for model-dependent $C_\theta$.

As can be seen when comparing figures \ref{constraints_dr7_ccov}  and \ref{constraints_dr7_vcov} the modeling error due to the approximation of constant $C$ is relatively small. Compared to the size of the $1\sigma$ intervals, the maximum likelihood positions are shifted by respectively $31\%$ for $\omega_m$ and $10\%$ for $\alpha$. The $1\sigma$ intervals also get reduced by respectively $31\%$ for $\omega_m$ and $29\%$ for $\alpha$. However we will see in section \ref{model_error_dr7simu} that the reduction of the $1\sigma$ region is not systematic.

\subsection{Quantifying the effect of $C_\theta$ on SDSS DR7-Full simulations}
\label{model_error_dr7simu}
The approximation of $C_\theta$ as a constant $C$ results in a modeling error, which potentially depends on the particular realization $\hat{\xi}$. So we want to quantify the general effect of this approximation on cosmological parameter constraints using a lot of realizations $\hat{\xi}$
\begin{equation}
\hat{\xi} \thicksim \mathcal{N}(\xi_{\theta_0},C_{\theta_0})
\label{model_simu}
\end{equation}
with the choice $\theta_0=(\omega_m, \alpha,B)=(0.13,1.0,2.5^2)$. For each realization $\hat{\xi}$ we compute the 2-dimensional posterior $p(\omega_m, \alpha \, | \, \hat{\xi})$ in the case of constant $C$ and model-dependent $C_\theta$. 

We look at two particular modeling errors
\begin{itemize}
\item Error on the position of the 1-dimensional posterior's maxima $\omega^\text{max}_m$ and $\alpha^\text{max}$
\item Error on the size of the $1\sigma$ intervals $\sigma_{\omega_m}$ and $\sigma_\alpha$
\end{itemize}

We adopt the following notations
\begin{eqnarray}
\label{delta_omega_max}
\delta \omega^\text{max}_m &=&  \left(\omega^{\text{max},C}_m -\omega^{\text{max},C_\theta}_m\right)/ \sigma^C_{\omega_m} \\ 
\label{delta_sigma_omega}
\delta \sigma_{\omega_m} &=& \left(\sigma^C_{\omega_m}-\sigma^{C_\theta}_{\omega_m} \right)/ \sigma^C_{\omega_m}  \\ 
\label{delta_alpha_max}
\delta \alpha^\text{max} &=& \left(\alpha^{\text{max},C} -\alpha^{\text{max},C_\theta} \right)/\sigma^C_\alpha \\ 
\label{delta_sigma_alpha}
\delta \sigma_\alpha &=&  \left(\sigma^C_\alpha-\sigma^{C_\theta}_\alpha \right)/\sigma^C_\alpha
\end{eqnarray}

We generate 2000 realizations following the model of formula (\ref{model_simu}) and look at the different quantities $\delta \omega^\text{max}_m$, $\delta \sigma_{\omega_m}$, $\delta \alpha^\text{max}$ and $\delta \sigma_\alpha$, which characterize the modeling error due to incorrect covariance matrix for each realization $\hat{\xi}$. Each quantity is divided by the $1\sigma$ interval size (the statistical uncertainty) in equations (\ref{delta_omega_max}), (\ref{delta_sigma_omega}), (\ref{delta_alpha_max}) and (\ref{delta_sigma_alpha}) in order to compare the modeling error to the statistical uncertainty. 

We compute the mean values $\langle \delta \omega^\text{max}_m \rangle$,$\langle \delta \sigma_{\omega_m} \rangle$, $\langle \delta \alpha^\text{max} \rangle$ and $\langle \delta \sigma_\alpha \rangle$ to investigate a systematic shift in the posterior's maxima or a systematic reduction of the $1\sigma$ intervals. However we found that these mean values are negligible compared to the $1\sigma$ interval sizes ($\approx 2\%$).

Next we compute the mean absolute values $\langle|\delta \omega^\text{max}_m| \rangle$,$\langle|\delta \sigma_{\omega_m}|\rangle$, $\langle|\delta \alpha^\text{max}|\rangle$ and $\langle|\delta \sigma_\alpha|\rangle$. $\langle |\delta \omega^\text{max}_m| \rangle$ and $\langle |\delta \alpha^\text{max}| \rangle$ give the mean modeling error on the position of the posterior's maxima compared to the $1\sigma$ interval sizes. On the other hand, $\langle |\delta \sigma_{\omega_m}| \rangle$ and $\langle |\delta \sigma_\alpha| \rangle$ give the mean modeling error on the $1\sigma$ interval sizes. These absolute values actually correspond to what is normally referred as the modeling error (indeed for a given realization $\hat{\xi}$, we do not really care about the sign of the error but only on its amplitude). We show our results in table \ref{modeling_error1}.

\begin{table}[htbp]
\caption{\label{modeling_error1}}
\begin{center}
\begin{tabular}{ccc} 
\tableskip\tableline\tableline\tableskip 
			      $\langle| \delta \omega^\text{max}_m |\rangle$ & $21\%$  \\
			      $\langle| \delta \sigma_{\omega_m} |\rangle$ &  $7.5\%$ \\
			      $\langle| \delta \alpha^\text{max} |\rangle$ &  $28\%$  \\
			      $\langle| \delta \sigma_\alpha |\rangle$ &  $10\%$   \\
\tableskip\tableline\tableline\tableskip
\end{tabular}
\end{center}
NOTES.---%
Importance of the modeling error compared to the $1\sigma$ interval size, both for the position of the posterior's maxima and for the size of $1\sigma$ intervals. We find a mean modeling error which is quite small compared to the $1\sigma$ interval sizes.
\end{table}

As shown in table \ref{modeling_error1}, there is a mean modeling error of $21\%$ to $28\%$ for the position of the posterior's maxima and $7.5\%$ to $10\%$ for the size of the $1\sigma$ intervals. So the position of the posterior's maxima is much more affected by the modeling error than the $1\sigma$ intervals. However the error stays quite small compared to the $1\sigma$ intervals. 

From table \ref{modeling_error1}, we see that the error on the extremities of the $1\sigma$ intervals is likely to stay below $21\%+7.5\%=28.5\%$ for $\omega_m$ and $28\%+10\%=38\%$ for $\alpha$. So a possible way to handle the modeling error (though it cannot be handled for sure, because it depends on the particular realization $\hat{\xi}$) is to multiply the size of $1\sigma$ intervals obtained with a constant covariance matrix $C$ by a factor $\approx1.3$ for $\omega_m$ and $\approx1.4$ for $\alpha$. In this way, the new $1\sigma$ intervals will very likely cover most of the real 1$\sigma$ intervals (i.e. the ones obtained with a model-dependent $C_\theta$).

\begin{figure}[h]
\plotone{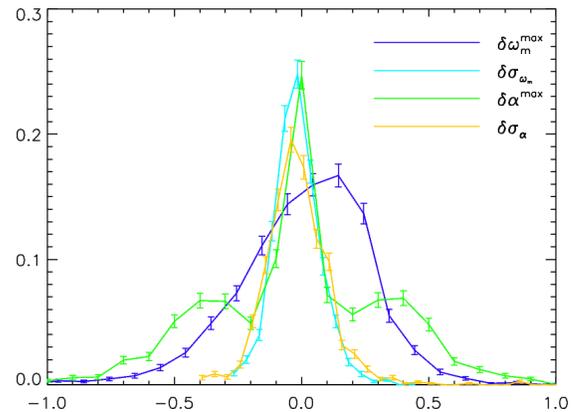} 
\caption{Estimated pdf of $\delta \omega^\text{max}_m$, $\delta \sigma_{\omega_m}$, $\delta \alpha^\text{max}$ and $\delta \sigma_\alpha$ using their histogram on 2000 realizations. Error bars give the Poisson uncertainty in the estimate due to finite number of realizations.} 
\label{modeling_error} 
\end{figure}

Let us illustrate more clearly how the modeling error can vary depending on the realization $\hat{\xi}$. On figure \ref{modeling_error} we show for each quantity $\delta \omega^\text{max}_m$, $\delta \sigma_{\omega_m}$, $\delta \alpha^\text{max}$ and $\delta \sigma_\alpha$ the estimated probability density function (pdf) from their histogram on 2000 realizations $\hat{\xi}$. We clearly see that the small modeling error varies depending on the realization $\hat{\xi}$.

Finally we perform a visual inspection of the 2-dimensional posteriors $p(\omega_m,D_V(0.3) \, | \, \hat{\xi})$ in both cases of constant covariance $C$ and model-dependent $C_\theta$. As in section \ref{model_error_dr7} we find deviations of the 2-dimensional posteriors compared to 2-dimensional Gaussians for most realizations $\hat{\xi}$. These deviations are located at high $\omega_m$ and low $\alpha$ and they happen both in the case of constant $C$ and model-dependent $C_\theta$. So they are simply due to the behavior of the model correlation function $\xi_\theta$ in this region. 

\subsection{Quantifying the effect of $C_\theta$ for next-generation surveys}
Finally we try to quantify this modeling error for next-generation surveys. Our procedure is simply to divide the covariance matrices $C$ and $C_\theta$ by a constant factor $c$ with $c=2$ and $c=4$, and repeat the analysis of section \ref{model_error_dr7simu}. To give an idea of what this represents in terms of survey size, we can approximate doubling the survey size as a factor $1/2$ in the covariance matrix

\begin{eqnarray}
C\left[ \frac{1}{2} \hat{\xi}_1 + \frac{1}{2} \hat{\xi}_2 \right] & \approx & \frac{1}{4} C\left[ \hat{\xi}_1\right]+\frac{1}{4} C\left[ \hat{\xi}_2 \right] \\
& \approx & \frac{1}{2} C\left[ \hat{\xi}_1\right]
\end{eqnarray}
because the estimated correlation function $\hat{\xi}_{12}$ of survey '1+2' is approximately the same as the mean of $\hat{\xi}_1$ and $\hat{\xi}_2$ for large enough surveys. So a factor $1/2$ in the covariance matrix is approximately equivalent to doubling the survey size, and similarly a factor $1/4$ in the covariance matrix is approximately equivalent to quadrupling the survey size.

Now we generate realizations from the model
\begin{equation}
\hat{\xi} \thicksim \mathcal{N}\left(\xi_{\theta_0},\frac{1}{c} \, C_{\theta_0}\right)
\label{model_simu2}
\end{equation}

The approximate likelihood (with constant covariance matrix) and real likelihood (with model-dependent covariance matrix) are now given by
\begin{eqnarray}
 \mathcal{L}^C_{\theta}(\hat{\xi}) & \propto &  e^{ -\frac{c}{2} \left\langle \hat{\xi}-\xi_{\theta}, C^{-1} (\hat{\xi}-\xi_{\theta}) \right\rangle} \\
 \mathcal{L}^{C_\theta}_{\theta}(\hat{\xi}) & \propto & |C_\theta|^{-1/2} e^{ -\frac{c}{2} \left\langle \hat{\xi}-\xi_{\theta}, C^{-1}_\theta (\hat{\xi}-\xi_{\theta}) \right\rangle} 
\end{eqnarray}

We repeat the analysis of table \ref{modeling_error1} with 2000 realizations of formula (\ref{model_simu2}) for each case $c=2$ and $c=4$. We show the results in table \ref{modeling_error2}.
\begin{table}[htbp]
\caption{\label{modeling_error2}}
\begin{center}
\begin{tabular}{ccccc} 
\tableskip\tableline\tableline\tableskip
									&  \multicolumn{1}{c}{$c=2$}& \multicolumn{1}{c}{$c=4$} \\
\tableskip\tableline\tableline\tableskip 
			      $\langle| \delta \omega^\text{max}_m |\rangle$ 	 & $16\%$ 	& $13\%$  \\
			      $\langle| \delta \sigma_{\omega_m} |\rangle$ 	 & $6.3\%$ 	& $5.1\%$ \\
			      $\langle| \delta \alpha^\text{max} |\rangle$		 &  $23\%$       	& $20\%$\\
			      $\langle| \delta \sigma_\alpha |\rangle	$	 	& $8.5\%$ 	& $6.9\%$   \\
\tableskip\tableline\tableline\tableskip
\end{tabular}
\end{center}
NOTES.---%
Importance of the modeling error compared to the $1\sigma$ intervals size, both for the position of the posterior's maxima and for the size of $1\sigma$ intervals when dividing $C$ and $C_\theta$ by factors $c=2$ and $c=4$. Again we find a mean modeling error which is quite small compared to the $1\sigma$ interval sizes. The error is smaller here than for the SDSS DR7-Full simulations, and it decreases with the survey size.
\end{table}

From table \ref{modeling_error2} we find again that there is mean modeling error which is quite small compared to the $1\sigma$ interval sizes. The modeling error mainly affects the position of the posterior's maxima. It is smaller here than for the SDSS DR7-Full simulations, and it decreases with the survey size. 

Our conclusion is that the approximation of $C_\theta$ as a constant $C$ only has a small impact on cosmological parameter constraints. As surveys get larger the modeling error decreases. Again an approximate way to handle this modeling error is to multiply the size of $1\sigma$ intervals by a factor $\approx1.3$. We emphasize that our study is not comprehensive and that we only took into account 3 parameters: $\theta=(\omega_m,\alpha,B)$.

This conclusion is a bit surprising since we found a strong dependence of $C_\theta$ on $\theta$. However it is easy to see that there is a competing effect at work in the likelihood. Let us remind the expression of the likelihood for a model-dependent covariance matrix $C_\theta$
\begin{equation}
 \mathcal{L}_\theta(\hat{\xi}) \propto  |C_\theta|^{-1/2} e^{ -\frac{1}{2} \left\langle \hat{\xi}-\xi_\theta, C^{-1}_\theta (\hat{\xi}-\xi_\theta) \right\rangle} 
\end{equation}
For example if we multiply the covariance matrix by a constant factor $c$ the terms $|C_\theta|^{-1/2}$ and $e^{ -\frac{1}{2} \left\langle \hat{\xi}-\xi_\theta, C^{-1}_\theta (\hat{\xi}-\xi_\theta) \right\rangle}$ will have competing effects, decreasing the overall effect on the likelihood. And we indeed verified that the term $|C_\theta|^{-1/2}$ has an important contribution in practice (i.e. if it is omitted, one obtains much greater changes of the likelihood contours).

Finally we also perform a visual inspection of the 2-dimensional posteriors $p(\omega_m,D_V(0.3) \, | \, \hat{\xi})$ in both cases of constant covariance $C$ and model-dependent $C_\theta$ for $c=2$ and $c=4$. Because the maximum likelihood is much closer to the real parameter $\theta_0$ of formula (\ref{model_simu2}) than in section \ref{model_error_dr7simu} (because variations of $\hat{\xi}$ are smaller), the region causing deviations to a 2-dimensional Gaussian is nearly always outside the 2 to 3$\sigma$ confidence region. So we find that realizations $\hat{\xi}$ of formula (\ref{model_simu2}) have 2-dimensional posteriors that can be very well approximated by 2-dimensional Gaussians.

\section{Conclusions}
\label{conclusion}
In this paper we have studied the influence of considering a realistic model-dependent covariance matrix $C_\theta$ instead of a constant covariance matrix $C$ of the estimated correlation function $\hat{\xi}$ for cosmological parameter constraints. The main difficulty comes from the very long computation time required to estimate such a model-dependent covariance matrix $C_\theta$. 

We have presented a new method to obtain a realistic model-dependent $C_\theta$ in a reasonable time, for a 3-dimensional parameter $\theta=(\omega_m,\alpha,b^2)$ using lognormal simulations. Compared to a constant covariance matrix, the computing time is multiplied by a factor roughly 20. We plan to release (as part of a general toolbox on the correlation function analysis of galaxy clustering) the different programs to estimate a model-dependent $C_\theta$ for different survey masks, selection functions and ranges of cosmological parameters.

Our first results concern the statistical modeling of the measured correlation function $\hat{\xi}$
\begin{equation}
\exists \, \theta \in \Theta \,\, \mbox{s.t.} \,\, \hat{\xi} \thicksim \mathcal{N}\left( \xi_\theta ,C_\theta \right) 
\end{equation}

We verified the absence of bias in our lognormal simulations, i.e. that the expected value of measured correlation function $\mathbb{E}(\hat{\xi})$ is indeed equal to the input model in our simulations $\xi_\theta$. Next we verified the Gaussianity of the measurement $\hat{\xi}$ using 80 Las Damas realizations, which are more realistic than our lognormal simulations. We estimated the probability density function of a $\chi^2$ statistic on these 80 realizations, and found that it is compatible with the expected result for Gaussian realizations. 

We also studied the dependence of $C_\theta$ with respect to $\omega_m$, $\alpha$ and $B=b^2$. We found that the effect of the amplitude parameter $b^2$ can be well approximated as a constant factor $b^4$ in the covariance matrix (for $b$ high enough, i.e. $>2$). For the two other parameters $\omega_m$ and $\alpha$, we found that their variations affect the whole shape of the covariance matrix. However $\omega_m$ has a bigger effect than $\alpha$ for usual ranges of parameter values.

Next we studied the implications of a model-dependent $C_\theta$ for cosmological parameter constraints. More precisely, we always compared the results obtained with $C_\theta$ to the results obtained with a constant $C=C_{\theta_0}$ for the particular value $\theta_0=(\omega_m,\alpha,b^2)=(0.13,1.0,2.5^2)$. 

For the SDSS DR7-Full sample, we obtained $\omega_m=0.145 \pm 0.016$ ($10.8\%$ precision) and $D_V(0.3)= 1104 \pm 105$ Mpc ($9.5\%$ precision) for a constant $C$, whereas we obtained  $\omega_m=0.140 \pm 0.011$ ($7.9\%$ precision) and $D_V(0.3)=1114 \pm 74$ Mpc ($6.7\%$ precision) for a model-dependent $C_\theta$. So there is only a small shift in the position of the posterior's maxima, and the $1\sigma$ intervals get a bit reduced when considering a model-dependent $C_\theta$. 

However this effect is not systematic and depends on the particular realization $\hat{\xi}$. In other words, approximating $C_\theta$ as a constant $C$ results in a modeling error both for the position of the posterior's maxima and for the size of the $1\sigma$ intervals, which depends on the particular realization $\hat{\xi}$. We quantified this modeling error using a lot of SDSS DR7-Full simulations
\begin{equation*}
\hat{\xi} \thicksim \mathcal{N}(\xi_{\theta_0},C_{\theta_0})
\end{equation*}

For each parameter, $\omega_m$ and $D_V(0.3)$, we studied the error in the position of the posterior's maximum and in the size of the $1\sigma$ interval. We found a mean modeling error in the position of the posterior's maxima approximately equal to $20\%$ to $30\%$ of the $1\sigma$ intervals. The error in the size of the $1\sigma$ intervals is smaller and is approximately equal to $10\%$.

Finally we did the same analysis for next-generation surveys, simply by dividing the covariance matrix by a factor $c$, with $c=2$ and $c=4$
\begin{equation*}
\hat{\xi} \thicksim \mathcal{N}\left(\xi_{\theta_0},\frac{1}{c} \, C_{\theta_0}\right)
\end{equation*}

We also found a small mean modeling error on the position of the posterior's maxima and on the size of the $1\sigma$ intervals. As the survey gets larger this modeling error decreases. More precisely if we multiply the size of the SDSS DR7-Full survey by a factor 4, the mean modeling error on the position of the posterior's maxima reaches $\approx 20\%$ of the $1\sigma$ interval size and the mean absolute error of the $1\sigma$ interval size reaches $\approx 6\%$.

So our conclusion is that the modeling error due to the approximation of $C_\theta$ as a constant $C$ is quite small. However for a safer analysis (though this modeling error cannot be handled for sure), one can multiply the size of $1\sigma$ intervals by a factor $\approx 1.3$.

This conclusion is a bit surprising since we found a strong dependence of $C_\theta$ on $\theta$. However there is a competing effect at work in the likelihood $\mathcal{L}_\theta(\xi)$ that tends to erase scaling effects.

Computing $C_\theta$ with a higher dimensional-parameter $\theta$ seems very difficult and cannot be addressed with our procedure yet. The approach proposed in \cite{Xu12} of a semi-analytic $C_\theta$ seems very promising in that respect. However it requires an ad hoc fitting of some parameters. In order to perform this parameter fitting, our simulations for a 3-dimensional parameter $\theta$ could be very interesting to use. Such an analysis would enable to see whether our conclusions are still correct when considering a full set of cosmological parameters.

\begin{acknowledgements}
Part of this work was supported by the European Research Council grant ERC-228261. We would like to thank the anonymous referee for helping to improve the quality of this paper.
      
Funding for the SDSS and SDSS-II has been provided by the Alfred P. Sloan Foundation, the Participating Institutions, the National Science Foundation, the U.S. Department of Energy, the National Aeronautics and Space Administration, the Japanese Monbukagakusho, the Max Planck Society, and the Higher Education Funding Council for England. The SDSS Web Site is http://www.sdss.org/.
      
\end{acknowledgements}

\appendix
\section{Optimal linear combination of estimators} 
\label{optimal_combination}

In this section we assume that we have two independent and unbiased Gaussian estimators $X_1$, $X_2$ (of dimension $n$) of $X_0$ with respective covariance matrices $C_1$ and $C_2$
\begin{eqnarray}
X_1 \thicksim \mathcal{N}(X_0,C_1) \\
X_2 \thicksim \mathcal{N}(X_0,C_2)
\end{eqnarray}

We consider an unbiased estimator $X$ of $X_0$ as a linear combination of $X_1$ and $X_2$
\begin{equation}
X=A X_1+ (Id-A) X_2
\end{equation}

with $A$ a square $n \times n$ matrix. The resulting covariance matrix is given by
\begin{equation}
\label{full_matrix}
C= \mathbb{E}[ X X^T]= A C_1 A^T + (Id-A) C_2 (Id-A)^T
\end{equation}

where we used the fact that $X_1$ and $X_2$ are independent. We will show that the following choice of $A$ gives an extremum of $\mbox{det} (C)$
\begin{eqnarray}
A & = & \left( C^{-1}_1+C^{-1}_2 \right)^{-1} C^{-1}_1 \\
\label{equation_A}
Id-A & = & \left( C^{-1}_1+C^{-1}_2 \right)^{-1} C^{-1}_2 
\label{equation_Id-A}
\end{eqnarray}

For this we use the following derivatives formulae, with $B$ a symmetric $n \times n$ matrix
\begin{eqnarray}
\frac{ \partial \mbox{det}(C)}{\partial C} & = & \mbox{det} (C) C^{-T} \\
\frac{ \partial (A B A^T)}{\partial A_{ji}} & = & A B J^{ij} + J^{ji} B A^T = A B J^{ij} + (A B J^{ij})^T
\end{eqnarray}

with $(J^{ij})_{kl}=\delta_{ik} \delta_{jl}$. Differentiating $\mbox{det} (C)$ with respect to $A$, we get

\begin{equation}
\frac{ \partial \mbox{det}(C)}{\partial A_{ji}} =  \sum_{kl} \mbox{det} (C) C^{-T}_{kl} \frac{ \partial C_{kl}}{\partial A_{ji}} 
\end{equation}

So it is sufficient to have for all $i,j$ that $\frac{ \partial C}{\partial A_{ji}}=0$ 
\begin{eqnarray}
\frac{ \partial C}{\partial A_{ji}} & = & \frac{ \partial (A C_1 A^T)}{\partial A_{ji}} + \frac{ \partial \left( (Id-A) C_2 (Id-A)^T \right)}{\partial A_{ji}} \\
& = & (A C_1- (Id-A) C_2) J^{ij} + [(A C_1 - (Id-A) C_2) J^{ij}]^T
\end{eqnarray}

Again it is sufficient to only have $A C_1- (Id-A) C_2=0$, which gives
\begin{eqnarray}
A (C_1+C_2) & = &C_2 \\
A & = &C_2 \left( C_1 + C_2 \right)^{-1} \\
A & = &C_2 C^{ -1}_2 \left( C^{-1}_1 + C^{ -1}_2 \right)^{-1} C^{-1}_1
\end{eqnarray}

So we obtain the solution given by equations (\ref{equation_A}) and equations (\ref{equation_Id-A})
\begin{eqnarray}
A & = & \left( C^{-1}_1+C^{-1}_2 \right)^{-1} C^{-1}_1 \\
Id-A & = & \left( C^{-1}_1+C^{-1}_2 \right)^{-1} C^{-1}_2 
\end{eqnarray}

Finally when using this expression of $A$ into equation (\ref{full_matrix}) we get
\begin{equation}
C= \mathbb{E}[ X X^T]= \left( C^{-1}_1+C^{-1}_2 \right)^{-1} 
\end{equation}

\end{document}